\documentclass[twocolumn]{aastex631} % Use APJ BibTeX style
\usepackage{CJKutf8}
\graphicspath{ {./figures} }
\usepackage{amsmath}

\newcommand{\CO}[2]{\mbox{$\mathrm{CO}\,(#1\text{--}#2)$}}

\begin{document}

%%%%%% Title %%%%%%
% Full titles can be a maximum of 100 characters, including spaces. 
% Title Format: Use title case, capitalizing the first letter of each word, except for certain small words, such as articles and short prepositions
\title{Multi-scale Gas Structure and Dynamics in an Extragalactic Central Molecular Zone}

\author[0009-0002-1791-4864]{Liam M. Wang}
\affiliation{Department of Astrophysical Sciences, Princeton University, 4 Ivy Lane, Princeton, NJ 08544, USA}
\author[0000-0003-0378-4667]{Jiayi~Sun \begin{CJK*}{UTF8}{gbsn}(孙嘉懿)\end{CJK*}}
\affiliation{Department of Physics and Astronomy, University of Kentucky, 506 Library Drive, Lexington, KY 40506, USA}
\affiliation{Department of Astrophysical Sciences, Princeton University, 4 Ivy Lane, Princeton, NJ 08544, USA}
\author[0000-0003-4209-1599]{Yu-Hsuan Teng}
\affiliation{Department of Astronomy, University of Maryland, 4296 Stadium Drive, College Park, MD 20742, USA}
\author[0000-0001-9605-780X]{Eric W. Koch}
\email{}
\affiliation{National Radio Astronomy Observatory, 800 Bradbury SE, Suite 235, Albuquerque, NM 87106, USA}
\author[0000-0002-6302-0485]{Sanghyuk Moon}
\affiliation{Korea Astronomy and Space Science Institute, 776 Daedeok-daero, Yuseong District, Daejeon, South Korea}
\affiliation{Department of Astrophysical Sciences, Princeton University, 4 Ivy Lane, Princeton, NJ 08544, USA}
\author[0000-0002-0119-1115]{Elias Oakes}
\affiliation{Department of Physics, University of Connecticut, 196A Auditorium Road, Storrs, CT 06269, USA}
\author[0000-0002-5204-2259]{Erik Rosolowsky}
\affiliation{Department of Physics, University of Alberta, Edmonton, AB T6G 2E1, Canada}

%%%%%% Spacing %%%%%%
% Use paragraph spacing of 1.5 or 2 (for double spacing, use command \doublespacing)

%%%%%% Abstract %%%%%%
%TC:ignore
\begin{abstract}
The structures and dynamics of the interstellar medium are governed by a combination of self-gravity, external gravity, and various sources of ordered and random motions on different spatial scales.
This paper uses ALMA CO\,(3--2) observations at $0\farcs1\approx5$~pc resolution to examine the scale dependence of molecular gas structure and dynamics in the central molecular zone (CMZ) of a nearby galaxy, NGC~3351.
We use the dendrogram technique to characterize hierarchical molecular gas structures spanning two decades in spatial scales and measure their size, gas mass, and velocity dispersion.
Their size--linewidth relation shows a power-law slope of 0.58, comparable to measurements for CMZs in other galaxies and suggestive of significant contribution from ordered motion on large scales.
We further decompose the observed velocity dispersion in each gas structure into ordered versus random motions.
The former appears stronger in gas structures at $\gtrsim30$~pc while the latter becomes more dominant at $\lesssim30$~pc.
Modulo uncertainties with the CO-to-H$_2$ conversion factor, the estimated gravitational free-fall time is comparable to the crossing time of ordered motions for structures on all spatial scales, and both becomes longer than the crossing time of random motions at small, $\lesssim10$~pc scales.
Our results highlight the varying sources and drivers of gas motions on different spatial scales in the CMZ of a Milky Way-like galaxy.
\end{abstract}
%TC:endignore

%%%%%% Main Text %%%%%%

\section{Introduction}

The interstellar medium (ISM) is shaped structurally and dynamically by a multitude of physical processes over a wide range of spatial scales \citep[see review by][]{McKee_Ostriker_2007}. For instance, the large-scale gravitational potential of a galaxy shapes its global ISM distribution and causes orbital and streaming motions in the gas. These ordered motions become subdominant on smaller scales relative to turbulence, which can create hierarchical substructures in the gas. In the densest substructures, the ISM self-gravity becomes important, under which the gas can collapse to form new stars.

Observational studies of the ISM structure and dynamics often rely on empirical scaling relations to probe this complex physics. Larson's relations connect the mass, size, and velocity dispersion of molecular clouds in the Solar Neighborhood and suggest turbulence and self-gravity as the dominant physics on individual molecular cloud scales \citep{Larson1981,Solomon1987}.
Later studies expanded these analyses to gas structures of similar or smaller sizes in other regions of our Galaxy \citep{HeyerDame2015} and in nearby galaxies \citep[e.g.,][]{Bolatto2008,FukuiKawamura2010,SchinnererLeroy2024}.
These studies show that the relative importance of galactic potential, self-gravity, turbulence, and ordered motions depends not only on the spatial scale of the gas structures in question, but also on properties of the host galaxy and location within the galaxy \citep[also see e.g.,][]{Oakes2025,Xie2025}.

Recently, an increasing number of studies have focused on gas properties close to galaxy centers, especially the gas-rich central molecular zones (CMZs) in barred spiral galaxies. Observations across dozens of such systems show evidence of elevated gas velocity dispersion relative to other types of galaxy centers but cannot pinpoint the nature of the gas motion \citep[e.g.,][]{Sun2018,Sun2020}.
Higher resolution observations of $\lesssim$10~pc of individual targets reveal strong system-to-system differences, some governed by galactic potential and orbital shear, others by turbulence, streaming motions, or even cloud-cloud collisions \citep[e.g.,][]{Shetty2012,Liu2021,Choi2023,Choi2024}.

This work aims to build on these previous efforts by analyzing high-resolution CO observations with the Atacama Large Millimeter/submillimeter Array (ALMA) targeting the CMZ of a nearby, Milky Way-mass, barred spiral galaxy, NGC~3351 \citep{Sun2024}.
Thanks to ALMA's exquisite capability and the proximity of NGC~3351 \citep[10~Mpc;][]{Anand2021}, this new dataset probes molecular gas structures over two decades in spatial scales (from a few pc to 100s of pc).
This wide dynamic range enables a detailed examination of the relative strengths of turbulence, ordered motions, and self-gravity as a function of spatial scale in a relatively face-on extragalactic CMZ.

\section{Data and Method}
\label{sec:datamethod}

\begin{figure*}[tb]
\centering    \includegraphics[width=1.0\textwidth]{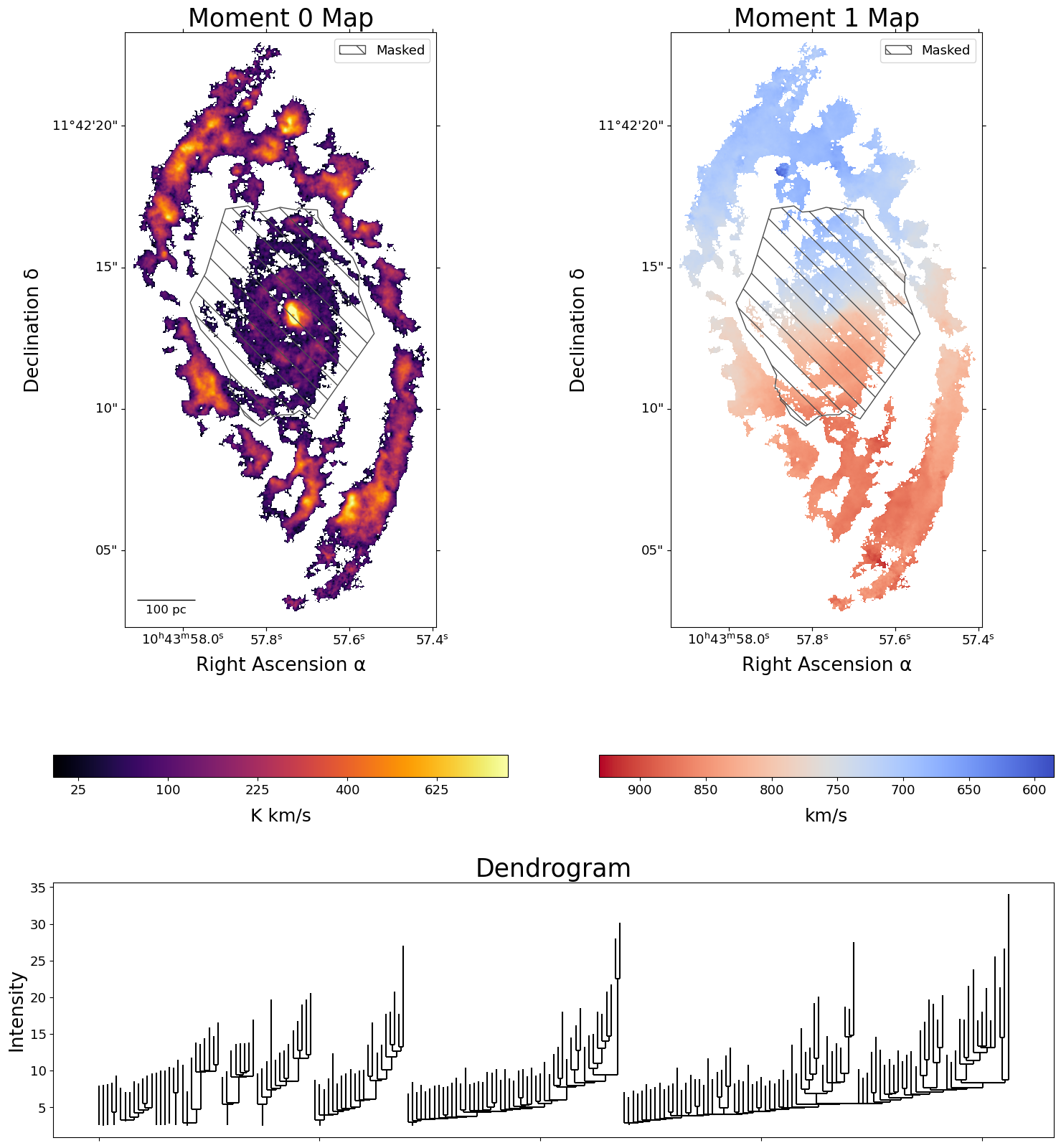}
\caption{\textit{(Top)} \CO32\ moment-0 and moment-1 maps for the CMZ of NGC~3351.
Our following structural analysis focuses on the outer star-forming ring (outside the hatched region; see \autoref{sec:datamethod}), where one sees the richest gas structure hierarchy.
\textit{(Bottom)} A dendrogram of the hierarchical gas structures across NGC~3351's star-forming ring. 
Vertical lines near the top of each structure tree represents local maxima in CO emission (a.k.a. ``leaves''), whereas those at the very bottom represents the largest connected structures in CO emission (a.k.a. ``trunks'').}
\label{fig:data}
\end{figure*}

We use ALMA \CO32\ data targeting the central $\sim$1~kpc$^2$ of NGC~3351 \citep{Sun2024}.
The CO data combines observations taken with multiple arrays and configurations to achieve both high resolution ($0.1''\sim5$~pc) and full recovery of flux on larger scales\footnote{We combine data from two 12m array configurations and the 7m array to recover flux on spatial scales up to the full size of the CMZ in NGC~3351.}.
The CO position-position-velocity (PPV) data cube used in this work has a velocity channel width of 2.5~km/s and a noise level of 1.3~K.
Interested readers are encouraged to consult \citet{Sun2024} for more details about the observational design and data reduction.

The top panels in \autoref{fig:data} shows the CO integrated intensity (``moment-0'') and centroid velocity (``moment-1'') maps.
CO emission is detected across the CMZ of NGC~3351, including an outer star-forming ring ($r\,{=}\,300{-}400$~pc), an inner circumnuclear disk ($r<200$~pc), and an innermost torus ($r<20$~pc).
The following analysis focuses on the gas along the outer ring, where one finds the richest hierarchy of substructures and signs of active star formation \citep[see][]{Calzetti2021,Sun2024}.
This is done by defining and applying a spatial mask that excludes all emission associated with or morphologically connected to the inner regions (\autoref{fig:data}).

To characterize the gas structure hierarchy, we create a dendrogram from the CO PPV data cube with \texttt{astrodendro} \citep{astrodendro}.
It highlights the nested organization of CO emission by identifying morphologically connected structures at all levels, ranging from the smallest isolated emission peaks, ``leaves'', to the largest connected structures in the field, ``trunks'' \citep{Rosolowsky2008}.
We carefully choose \texttt{astrodendro} input parameters to ensure that only significant CO emission structures are retained in the dendrogram.
Specifically, we use \texttt{astrodendro} to consider all CO emission within a custom-made signal mask, which we create following the ``strict'' masking scheme\footnote{Specifically, we include CO detection above 2$\sigma$ over 4 consecutive channels, and then expand to all morphologically connected CO emission above 2$\sigma$ over 2 channels in the cube.} defined by \citet{Leroy2021}.
We also require that all identified structures should be no smaller than the beam, and that each ``child'' structure should be brighter than its ``parent'' structure by at least twice the rms noise.

We show the resulting dendrogram in the bottom panel of \autoref{fig:data}.
In total, the dendrogram includes 398 unique gas structures (``branches'' and ``leaves'') represented by vertical lines.
These structures are organized in a tree-like way and can be traced down to 16 unique ``trunks'' as the largest morphologically connected CO emission in the data.
The highest ``leaves'' in the dendrogram instead correspond to the brightest structures visible in the moment-0 map.

We additionally calculate a few key physical properties for all gas structures (all ``branches'' and ``leaves'') in the dendrogram, as detailed in the following subsections.

\subsection{Structure Size}
\label{sec:datamethod:radius}

\texttt{Astrodendro} reports a radius (in units of arcsec) for each structure based on the second moment of its on-sky emission distribution. We convert this angular size into a physical size, $r_\mathrm{phy}$, according to the distance to NGC~3351.

\subsection{Gas Mass}
\label{sec:datamethod:mass}

We use the \CO32\ luminosity reported by \texttt{astrodendro} (which sums over the observed flux density of all voxels within a structure)
and convert it into molecular gas mass via
\begin{align}
M_\mathrm{mol} &= \alpha_\mathrm{CO(3{-}2)} L_\mathrm{CO(3{-}2)}~. \label{eq:mass1}
\end{align}
\noindent Here $\alpha_\mathrm{CO(3{-}2)}$ is the $\text{CO(3--2)-to-H}_2$ conversion factor at the location of each structure.
To obtain reliable $\alpha_\mathrm{CO(3{-}2)}$ estimates, we derive a $\alpha_\mathrm{CO(3{-}2)}$ map at $\sim$100 pc resolution, using all the CO isotopologue data in \citet{Teng2022} and following their procedure for measuring $\alpha_\mathrm{CO}$ based on Large Velocity Gradient (LVG) modeling \citep[see also][Appendix B]{Teng2023}. 
First, we convolve our \CO32\ data to match the minimum common beam size of $2.1''$ ($\sim$100 pc for NGC~3351) as constrained by the \CO10 data resolution in \citet{Teng2022}. We then incorporate the convolved \CO32 data into their one-component LVG modeling and Bayesian likelihood analysis. Next, by replacing $I_\mathrm{CO(1-0)}$ with $I_\mathrm{CO(3-2)}$ in \citet[Equation~10]{Teng2022}, we obtain pixel-by-pixel probability distributions of $\alpha_\mathrm{CO(3{-}2)}$ in the CMZ of NGC~3351, all of which are robustly constrained by seven CO isotopologue line measurements including our \CO32. Finally, we extract the median value of each probability distribution to be our $\alpha_\mathrm{CO(3{-}2)}$ solutions.

The derived $\alpha_\mathrm{CO(3{-}2)}$ values are typically $0.3{-}0.7$ $\rm M_\odot\,pc^{-2}\,(K\,km\,s^{-1})^{-1}$ in our area of interest.
Estimated via multi-line LVG modeling, these values reflect variations in the modeled gas density and temperature on $\gtrsim$100~pc scale (i.e., the resolution of the observations used in \citealt{Teng2022}).
However, we note that the gas conditions likely also vary on scales smaller than 100~pc.
Given the non-linear nature of LVG modeling, our low-resolution $\alpha_\mathrm{CO(3{-}2)}$ estimate might not accurately recover the true average value, therefore potentially introducing additional bias and/or scatter on the gas mass estimates (see further discussion in \autoref{sec:results:timescales}).

\subsection{Free-Fall Time}
\label{sec:datamethod:tff}

We calculate the gravitational free-fall time for each structure:
\begin{equation}
t_{\text{ff}} = \sqrt{\frac{3 \pi}{32 G \rho}} = \sqrt{\frac{\pi^2 r^3_\mathrm{phy}}{8 G M_\mathrm{mol}}}~.
\label{eq:tff}
\end{equation}
\noindent Here we assume spherical symmetry and use the gas mass and physical radius derived in \autoref{sec:datamethod:radius}--\ref{sec:datamethod:mass}. 

\subsection{Velocity Dispersions}
\label{sec:datamethod:vdisp}

\texttt{astrodendro} also reports a total velocity dispersion, $\sigma_\mathrm{v}$, for each structure in the dendrogram.
This $\sigma_\mathrm{v}$ includes all gas motions relative to the velocity centroid of the structure without distinguishing the nature of such motions.
We attempt to distinguish two general types of gas motion and quantify their contributions to the total $\sigma_\mathrm{v}$: (1) \textit{ordered motions} such as differential galactic rotation and spin/rotation of a gas structure; and (2) \textit{random motions} such as small-scale turbulence within a gas structure.
Using two different methods for separating ordered and random motions, we assess the level of systematic uncertainties associated with this process.

The first method for separating ordered and random motions models the former with a simple linear velocity gradient across each structure \citep[following][among other works]{Goodman1993, Liu2021}.
That is, we define a velocity gradient model for each structure as:
\begin{equation}
v_{\text{mod}}(\alpha, \delta) = v_0 + \omega_\alpha (\alpha - \alpha_{\text{ctr}}) \cos(\delta_{\text{ctr}}) + \omega_\delta (\delta - \delta_{\text{ctr}})~.
\label{eq:vmodel}
\end{equation}
\noindent Here $v_0$ is the systemic velocity at the structure center $(\alpha_\mathrm{ctr},\;\delta_\mathrm{ctr})$, and $\omega_\alpha$ and $\omega_\delta$ are the best-fit velocity gradient components along the RA and Dec direction, respectively.
For the gas structures studied here, we measure velocity gradients on the order of 0.2--0.6 km/s/pc, which are larger than the typical values found for molecular clouds in the Milky Way disk \citep[e.g.,][]{Imara2011} and more comparable to those found in the centers of nearby galaxies \citep[e.g.,][]{Liu2021}.

\begin{figure*}[htbp]
\centering
\includegraphics[width=0.95\textwidth]{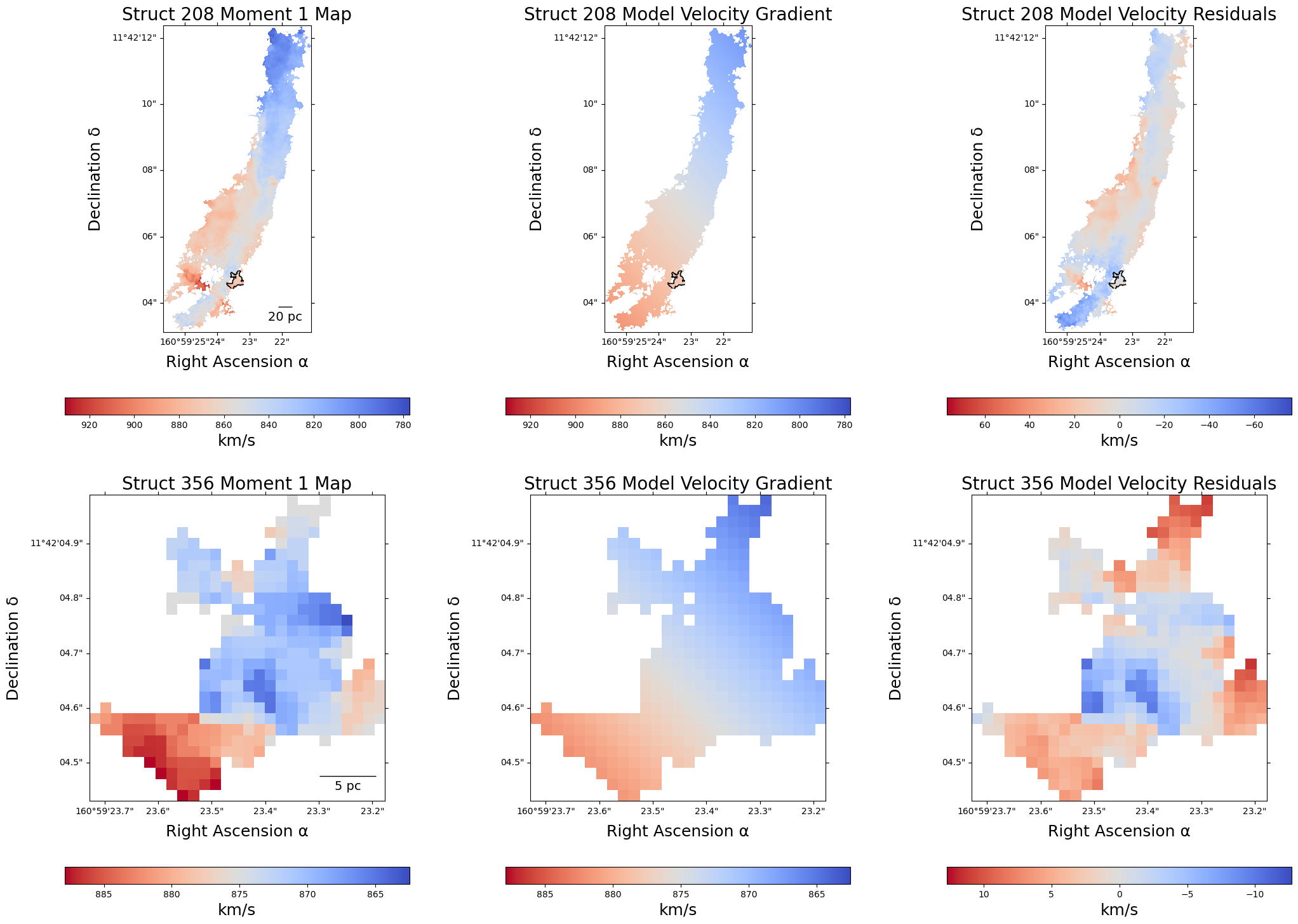}
\caption{\CO32\ moment-1 maps \textit{(left)}, model velocity gradient maps \textit{(middle)}, and velocity residual maps \textit{(right)} for two identified structures in the dendrogram.
The velocity gradient model and its residual provides one way to distinguish ordered versus random motions within each structure (see \autoref{sec:datamethod:vdisp}).
The top row shows a large trunk (structure \#208, $r_\mathrm{phy}=47$~pc) located on the southwest side of NGC~3351's star-forming ring (see \autoref{fig:data}).
This structure appears highly elongated and with complex, ordered velocity structures, which is not representative of most structures studied in this work.
The bottom row shows a smaller leaf (structure \#365, $r_\mathrm{phy}=5$~pc) inside structure \#208 (corresponding to the black contour in the top panels).}
\label{fig:struct}
\end{figure*}

We then fit our velocity gradient model for each structure in PPV space, weighted by the brightness temperature at each pixel.
A demonstration of this process for two connected structures (\#208 and \#356) in the dendrogram is shown in \autoref{fig:struct}.

Based on the best-fit velocity gradient model for each structure, we determine the \textit{random motion} velocity dispersion from the rms residual about the best-fit model:
\begin{equation}
\sigma_{\text{rand}} = \sqrt{\frac{\sum\limits_{\alpha, \delta, v}T_{\alpha, \delta, v} \left[v - v_{\text{mod}}(\alpha, \delta)\right]^2}{\sum\limits_{\alpha, \delta, v} T_{\alpha, \delta, v}}}~.
\label{eq:rand_hi}
\end{equation}
\noindent Here $T_{\alpha, \delta, v}$ represents the CO brightness temperature at any given spatial location and velocity channel in PPV space.
We then determine the corresponding velocity dispersion for the ordered motion from the quadrature difference between the total velocity dispersion and the random motion component:
\begin{equation}
\sigma_{\text{ord}} = \sqrt{\sigma_\mathrm{v}^2 - \sigma_{\text{rand}}^2}.
\label{eq:ord}
\end{equation}

The linear velocity gradient model (\autoref{eq:vmodel}) is a very restrained representation of ordered motions in a realistic gas structure.
The true ordered motion in the gas---which can include differential galactic rotation, streaming motions, and gas cloud rotation or spin---is often more complicated than a simple velocity gradient.
This is especially a concern for many of the larger ``branches'' and ``trunks'' in the dendrogram, whose internal kinematics are well resolved by our CO observations.
Thus our calculations from \autoref{eq:rand_hi}--\ref{eq:ord} would likely underestimate $\sigma_\mathrm{ord}$ and overestimate $\sigma_\mathrm{rand}$.

To address this issue, we introduce a second method for separating random and ordered motions.
Instead of the simple velocity gradient model, we use the observed centroid velocity of the structure (i.e., moment-1) at each spatial location as an alternative model for the ordered motion.
We then use the residual velocity dispersion about this new model as our alternative estimate for the random motion:
\begin{equation}
\sigma_{\text{rand}} = \sqrt{\frac{\sum\limits_{\alpha, \delta, v}T_{\alpha, \delta, v}[v - v_\mathrm{mom1}(\alpha, \delta)]^2}{\sum\limits_{\alpha, \delta, v}T_{\alpha, \delta, v}}}~.
\label{eq:rand_lo}
\end{equation}
The corresponding ordered motion estimate is again defined as the quadrature difference between $\sigma_\mathrm{v}$ and this alternative $\sigma_\mathrm{rand}$ per \autoref{eq:ord}.

This second method would likely overestimate the true ordered motions and underestimate the random motions, especially for highly resolved structures.
This is because random motions like turbulence can also create variations in the centroid velocity from place to place on intermediate scales, which our estimate for random motions with \autoref{eq:rand_lo} would fail to capture.

In the following analysis, we will treat the $\sigma_\mathrm{rand}$ estimates from \autoref{eq:rand_hi} and \autoref{eq:rand_lo} as upper and lower limits on the true random motion velocity dispersion.
The corresponding $\sigma_\mathrm{ord}$ estimates would serve as lower and upper limits for the true ordered motion, respectively.

\subsection{Crossing Times}
\label{sec:datamethod:tcr}

Lastly, we calculate the crossing time of various types of motions for each structure.
For instance, the overall crossing time is defined as the ratio between the structure radius and the total velocity dispersion:
\begin{equation}
t_\mathrm{cr} = \frac{r_\mathrm{phy}}{\sigma_\mathrm{v}}~.
\label{eq:t_cr}
\end{equation}
\noindent We replace $\sigma_\mathrm{v}$ in the above equation with $\sigma_\mathrm{rand}$ and $\sigma_\mathrm{ord}$ when calculating the crossing time for the random and ordered motions, respectively.

\newpage
\section{Results}
\label{sec:results}

\begin{figure*}[tb!]
\centering
\includegraphics[width=1.0\textwidth]{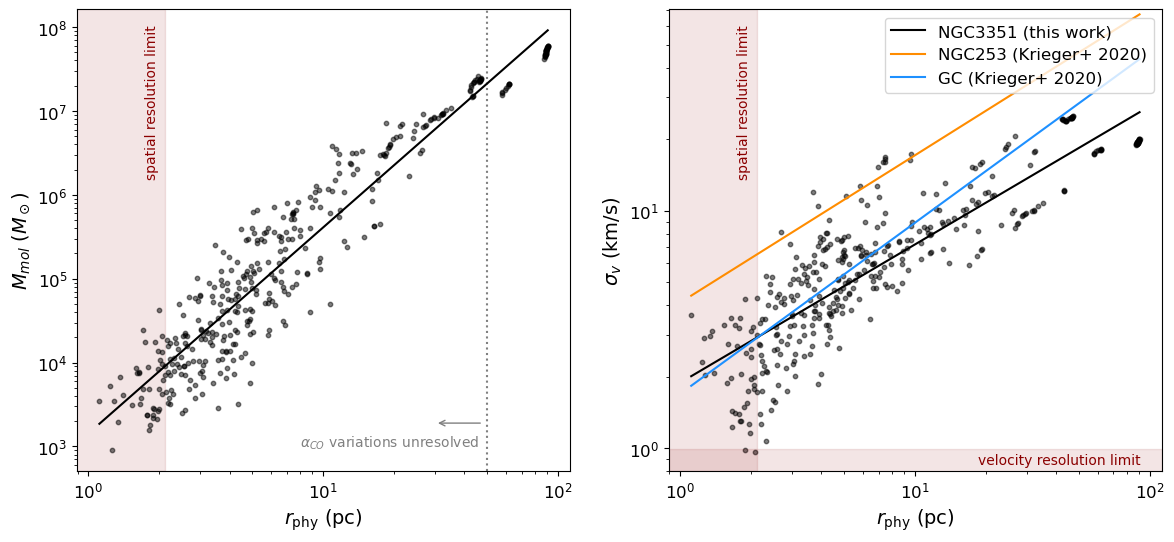}
\caption{\textit{(Left)} The size--mass relation among all identified gas structures, with the best-fit power law relation $M_\mathrm{mol} \propto r_\mathrm{phy}^{2.54}$ shown by a black line.
The vertical shaded region shows where the measured size becomes unreliable due to spatial resolution limit.
Besides, the gas mass measurements for smaller structures may also be inaccurate (gray dotted line) due to the coarser spatial resolution of the $\alpha_\mathrm{CO(3{-}2)}$ constraints \citep{Teng2023}.
\textit{(Right)}
The corresponding size--linewidth relation.
The best-fit power law $\sigma_\mathrm{v} \propto r_\mathrm{phy}^{0.58}$ has a similar slope but a lower intercept than that found for the center of NGC~253 \citep{Krieger2020}.
The intercept is closer to that found for the Milky Way's CMZ.
In addition to the same spatial resolution limit as shown in the left panel, the horizontal shaded region marks where the measured $\sigma_\mathrm{v}$ becomes unreliable due to finite velocity resolution.}
\label{fig:larson}
\end{figure*}

Our analysis covers the 398 unique gas structures across the star-forming ring in the CMZ of NGC~3351.
These structures span nearly two decades in their physical sizes, from a few parsecs to $\sim$100~pc.
Such a large range of structure sizes is rarely achieved in extragalactic observations \citep[c.f.][]{Oakes2025} due to finite resolution, sensitivity, and/or the use of ``cloud-finding'' algorithms that often pick up marginally resolved structures.
Here we take advantage of this spatial scale baseline and examine the scale-dependence of structure mass and velocity dispersions, as well as the relative strengths of ordered motion, random motion, and self-gravity.

\subsection{Larson's Relations}
\label{sec:results:scaling}

We show the size--mass relation and the size--linewidth relation among all gas structures in \autoref{fig:larson}.
The former spans an extensive range in gas mass from ${\sim}10^3\rm\,M_\odot$ to ${\sim}10^8\rm\,M_\odot$.
It is well described by a power law fit found using a linear \texttt{np.polyfit} fit on a power law dependence:
\begin{equation}
\frac{M_\mathrm{mol}}{10^5\rm\,M_\odot} = (4.1 \pm 0.4)\left(\frac{r_\mathrm{phy}}{10\rm\,pc}\right)^{2.54 \pm 0.04}~.
\end{equation}
At face value, this slope of 2.54 suggests that larger structures tend to have higher surface densities and lower volume densities.
However, it is important to note that the slope depends critically on the treatment of $\alpha_\mathrm{CO(3{-}2)}$ for structures of various sizes.
Due to the limited spatial resolution of the $\alpha_\mathrm{CO(3{-}2)}$ constraints from \citet{Teng2023}, our gas mass estimates for most structures smaller than $\sim50$~pc may not be totally accurate, and this would also affect the best-fit power law slope.
The size--linewidth relation among all identified gas structures is also reasonably described by a power law:
\begin{equation}
\frac{\sigma_\mathrm{v}}{\rm\,km\,s^{-1}} = (7.2 \pm 0.3)\left(\frac{r_\mathrm{phy}}{10\rm\,pc}\right)^{0.58 \pm 0.02}~.
\end{equation}
\autoref{fig:larson} compares this relation with those found with \CO32\ data for the CMZs of NGC~253 and our Milky Way\footnote{We note that these literature measurements are derived for the entire CMZs of NGC~253 and the Milky Way without excluding the circumnuclear molecular gas.} \citep[as reported in][]{Krieger2020}.
We find a similar power law slope as that reported for NGC~253, but the normalization is less than half the NGC~253 relation \citep[$\sigma_\mathrm{v,\,10pc}=17\rm\;km/s$;][]{Krieger2020}.
The comparison between the CMZs in NGC~3351 and the Milky Way galactic center (GC) instead shows a similar normalization ($\sigma_\mathrm{v,\,10pc}=8.9\rm\;km/s$ for the latter), though the slope may be somewhat steeper in the Milky Way GC \citep[$\sim$0.72;][]{Krieger2020}.
These differences are attributable to NGC~253 having a more compact and extreme CMZ than NGC~3351 or the Milky Way, with on average higher gas surface densities and correspondingly larger gas velocity dispersion at a given physical scale. 

The $\sigma_\mathrm{v}$ shown in \autoref{fig:larson} right panel is the total velocity dispersion, which includes various sources of ordered and random motions.
We expect the contribution from ordered motion to be more pronounced near galaxy centers, where galactic rotation and streaming motions often have stronger influence on the gas kinematics especially on larger scales.
Some recent observations find steeper size-linewidth relations in galaxy centers \citep[slope $\approx0.6{-}0.8$; e.g.,][]{Xie2025} compared to the Solar neighborhood \citep[slope $\approx0.5$;][]{Larson1981,Solomon1987}.
This can be attributed to gas motions being affected more by the external galactic potential in galaxy centers \citep[e.g.,][]{Mie2018}.

In the next subsection, we explore the scale-dependence of ordered and random motions with our quantitative measurements of these components further.

\subsection{Ordered versus Random Motions}
\label{sec:results:ordrand}

\begin{figure}[htbp]
\centering
\includegraphics[width=1.0\columnwidth]{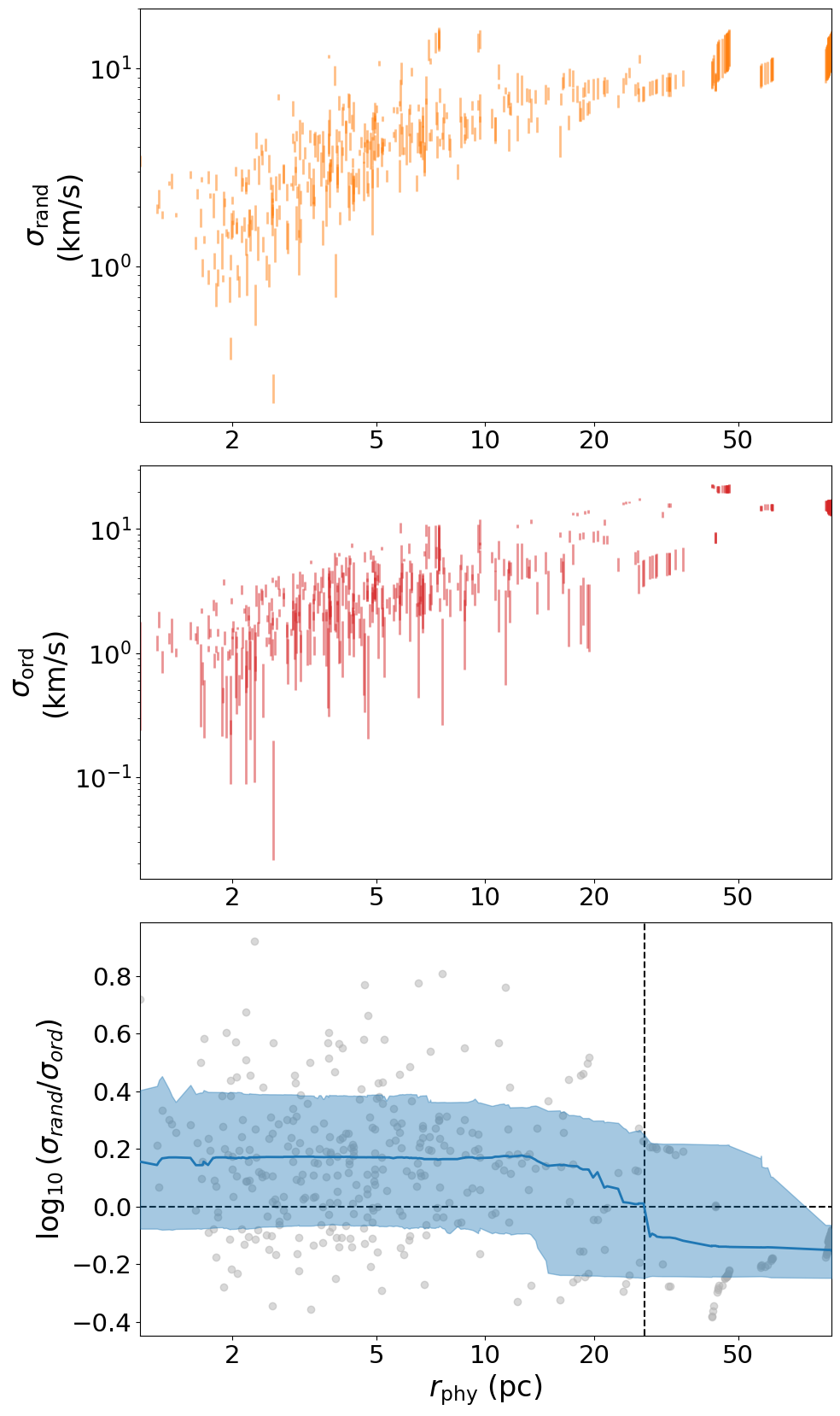}
\caption{\textit{(Top)} Estimated velocity dispersion of random motions as a function of structure size. The extent of each vertical bar represents the upper and lower limits on $\sigma_\mathrm{rand}$ (see \autoref{sec:datamethod:vdisp}).
\textit{(Middle)} Similar to the top panel, but shows the effective velocity dispersion of ordered motion, $\sigma_\mathrm{ord}$.
\textit{(Bottom)} Ratio between $\sigma_\mathrm{rand}$ and $\sigma_\mathrm{ord}$ (in logarithmic scale) as a function of structure size.
The gray data points mark the geometric mean between the upper and lower limits on this ratio.
Overlaid on top of these data points is the running median curve and the 16--84th percentile range. 
While both velocity dispersion components increase with spatial scale, random motion appears stronger than ordered motion for most structures at $\lesssim30$~pc (see the vertical dashed line where the running median curve first crosses zero); the opposite is true only for the largest structures.}
\label{fig:vdisp}
\end{figure}

We show the velocity dispersions of random and ordered motions versus structure size in the top and middle panels of \autoref{fig:vdisp}.
These velocity dispersions both span a similar range, from $\lesssim1\rm\,km/s$ on a few parsecs scale to $\gtrsim10\rm\,km/s$ on $\sim100$~pc scale.
Comparing the upper and lower limits derived from \autoref{eq:rand_hi}--\ref{eq:rand_lo}, we see that the bounds typically differ from each other by ${\sim}\,0.05$~dex for $\sigma_\mathrm{rand}$ and ${\sim}\,0.13$~dex for $\sigma_\mathrm{ord}$ (i.e., lengths of the vertical lines in \autoref{fig:vdisp} top and middle panels). 
While these uncertainties make it challenging to model the quantitative relationship of either component with spatial scale, it is clear that both components increase systematically with spatial scale. 

We further show the typical ratio of $\sigma_\mathrm{rand}/\sigma_\mathrm{ord}$ versus spatial scale in the bottom panel of \autoref{fig:vdisp}.
Focusing on the distribution of this random-to-ordered motion ratio at various spatial scales (i.e., running median plot across 10 $r_\mathrm{phys}$ bins), we find that most structures on $\lesssim30$~pc scales tend to have larger random motion than ordered motion.
The latter seems to dominate the total velocity dispersion budget only in the largest structures.
This general trend is what one would expect for ordered motions driven by large-scale processes, such as galactic rotation (see \autoref{sec:discuss}) and/or streaming motions driven by stellar bars and other non-axisymmetric galactic structures.
It is also consistent with our interpretation of the steep size--linewidth relation earlier in \autoref{sec:results:scaling}.
Our results broadly agree with those reported by \citet{Xie2025}, though they focused on molecular clouds in the Solar Neighborhood and found a slightly larger transition scale ($\sim100$~pc).

\subsection{Timescale Comparisons}
\label{sec:results:timescales}

\begin{figure*}[htbp]
\centering
\includegraphics[width=1.0\linewidth]{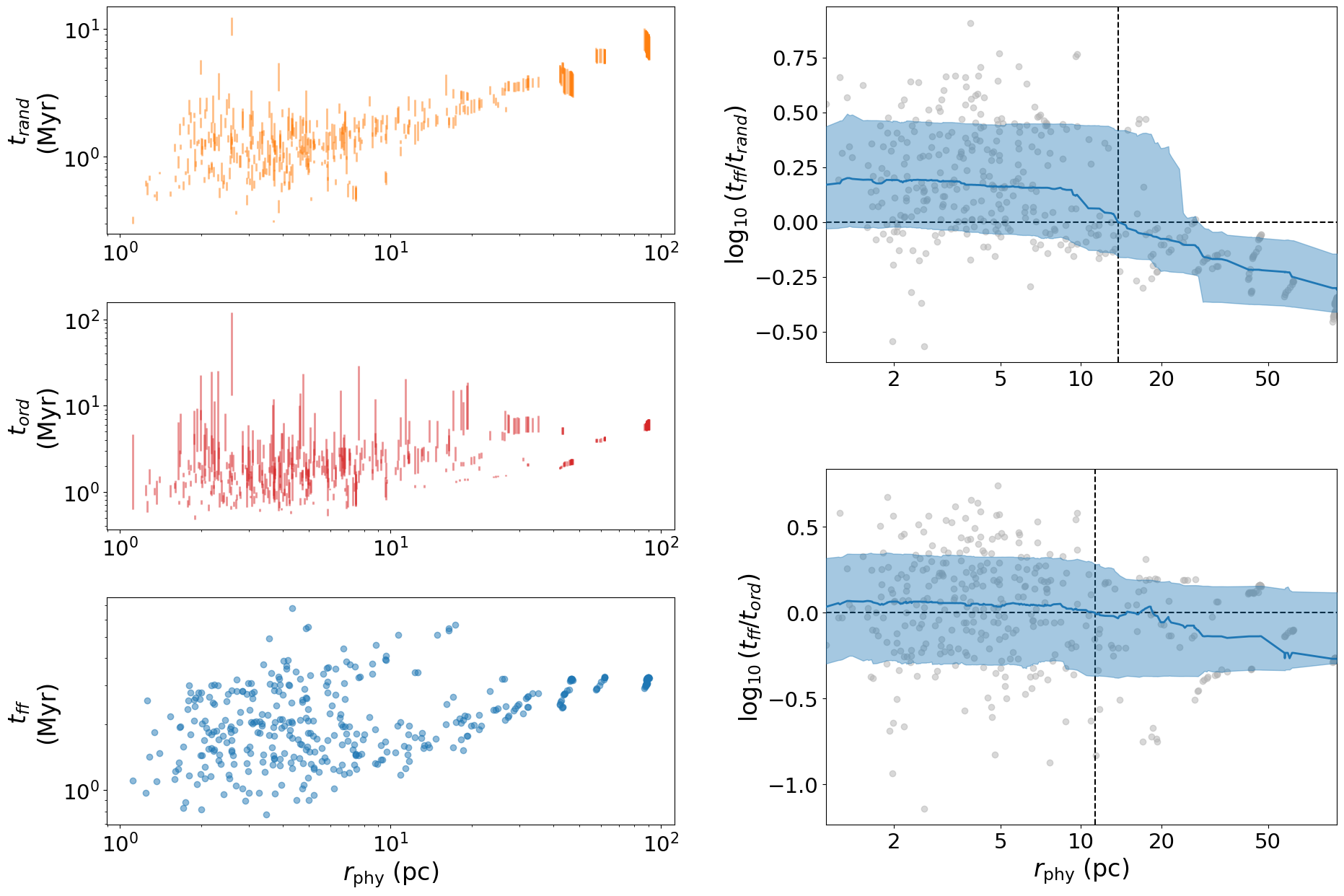}
\caption{\textit{(Left)} Random motion crossing time (orange), ordered motion crossing time (red), and gravitational free-fall time (blue) as functions of structure size.
As in \autoref{fig:vdisp}, we use vertical lines to show the range of $t_\mathrm{cr,\,rand}$ and $t_\mathrm{cr,\,ord}$ allowed between the upper and lower limits.
All three timescales tend to increase with structure sizes, although there are considerable amounts of scatter especially among structures at $\lesssim10$~pc.
\textit{(Right)} Ratios of free-fall time over the two crossing timescales (in logarithmic scale) as functions of structure size. $t_\mathrm{ff}$ tends to be longer than $t_\mathrm{cr,\,rand}$ on small scales ($\lesssim10$~pc) and shorter on large scales.
In contrast, $t_\mathrm{ff}$ appears on par with $t_\mathrm{cr,\,ord}$ for gas structures at all scales probed by our data.}
\label{fig:timescale}
\end{figure*}

We also compare the effects of ordered and random motions to that of self-gravity in gas structures on various spatial scales.
We do this by examining the scale-dependence of the corresponding physical timescales: the crossing time of ordered motion, $t_\mathrm{cr,\,ord}$, and of random motion, $t_\mathrm{cr,\,rand}$, and the gravitational free-fall time, $t_\mathrm{ff}$ (see \autoref{sec:datamethod:tff} and \ref{sec:datamethod:tcr}).
This is shown in the left panels in \autoref{fig:timescale}.

The crossing times and the free-fall time all fall in a similar range of ${\sim}\;1{-}10$~Myr.
Almost all structures larger than ${\sim}\;10$~pc show long crossing times and free-fall times (${\gtrsim}\;3$~Myr); smaller structures instead display a wide range of timescales without clear trends.
There also seems to be a distinct ``upper branch'' of the $t_\mathrm{ff}$ population, which mostly corresponds to small structures ($\lesssim20~\mathrm{pc}$) around the outermost portions of the CMZ ring, where the overall gas density is lower and therefore $t_\mathrm{ff}$ is higher.

It is interesting to compare these timescale estimates with literature measurements for gas structures in different systems.
For example, \citet{Sun2022} report typical turbulence crossing time of ${\sim}\;10{-}30$~Myr and free-fall time of ${\sim}\;5{-}15$~Myr for a large sample of extragalactic molecular clouds at 150~pc uniform resolution.
Detailed studies of individual galaxies with gas-rich CMZs \citep[e.g., NGC~253;][]{Leroy2015} tend to find shorter timescales that are more consistent with our results.
These results suggest a more rapid evolution of gas structures in CMZs than in the outer disks of local star-forming galaxies.

We then compare the three estimated timescales in our work by examining their ratios as functions of spatial scale in the right panels of \autoref{fig:timescale}.
For large structures $\gtrsim10$~pc, we find that $t_\mathrm{ff}$ tend to be shorter than $t_\mathrm{cr,\,rand}$ but generally comparable to $t_\mathrm{cr,\,ord}$.
This suggests that ordered motion (but not random motion) can counteract self-gravity in these larger gas structures, again consistent with the notion that the large-scale gas dynamics is dominated by ordered motions.

For smaller structures, we find that $t_\mathrm{ff}$ is still on par with $t_\mathrm{cr,\,ord}$ but now generally longer than $t_\mathrm{cr,\,rand}$. More quantitatively, $43\%$ of structures under $10$~pc have $t_\mathrm{ff}$ shorter than $t_\mathrm{cr,\,ord}$, but only $21\%$ have $t_\mathrm{ff}$ shorter than $t_\mathrm{cr,\,rand}$.
This suggests that most small gas structures have more than enough random motions to counteract self-gravity and prevent collapse and star formation.
While that relationship is possible for a fraction of small gas clumps not dense enough to form stars, it is unlikely to be true for \textit{all} small gas clumps, because star formation takes place at various regions along NGC~3351's nuclear ring \citep[e.g.,][]{Calzetti2021,Sun2024}.
We note that the uncertain $\alpha_\mathrm{CO(3{-}2)}$ at small spatial scales (see \autoref{sec:datamethod:mass}) becomes a more serious concern here.
As \CO32\ becomes increasingly optically thick in smaller, denser gas clumps, we expect to underestimate their gas mass based on the large-scale average $\alpha_\mathrm{CO(3{-}2)}$.
In this case, the true volume density should be systematically higher and the free-fall time shorter on small scales than our current estimates.

The timescale ratios examined in \autoref{fig:timescale} are closely related to the virial parameter, which appears frequently in the molecular cloud literature \citep[e.g.,][]{Liu2021,Oakes2025}.
Briefly, in the scenario where ordered motion is negligible and random motion provides an effective pressure support, the dynamical evolution of a gas structure can be determined from its virial parameter, $\alpha_\mathrm{vir}\equiv-2\mathcal{T}/\mathcal{W}$, where $\mathcal{T}$ and $\mathcal{W}$ are the kinetic and gravitational potential energy.
For structures of a certain geometry and density profile, the virial parameter is proportional to the squared timescale ratio $\alpha_\mathrm{vir} \propto (t_\mathrm{ff}/t_\mathrm{cr,\,rand})^2$.
However, in cases where ordered motions such as shear and rotation are present, the dynamical balance of a gas structure cannot be determined with the same simple $\alpha_\mathrm{vir}$ calculation---at least not without substantial revision of its definition \citep[see][]{Liu2021}.
Such a thorough virial analysis is beyond the scope of this work but an ideal topic for future study.

\section{Discussion}
\label{sec:discuss}

\begin{figure*}[htbp]
\centering
\includegraphics[width=1.0\linewidth]{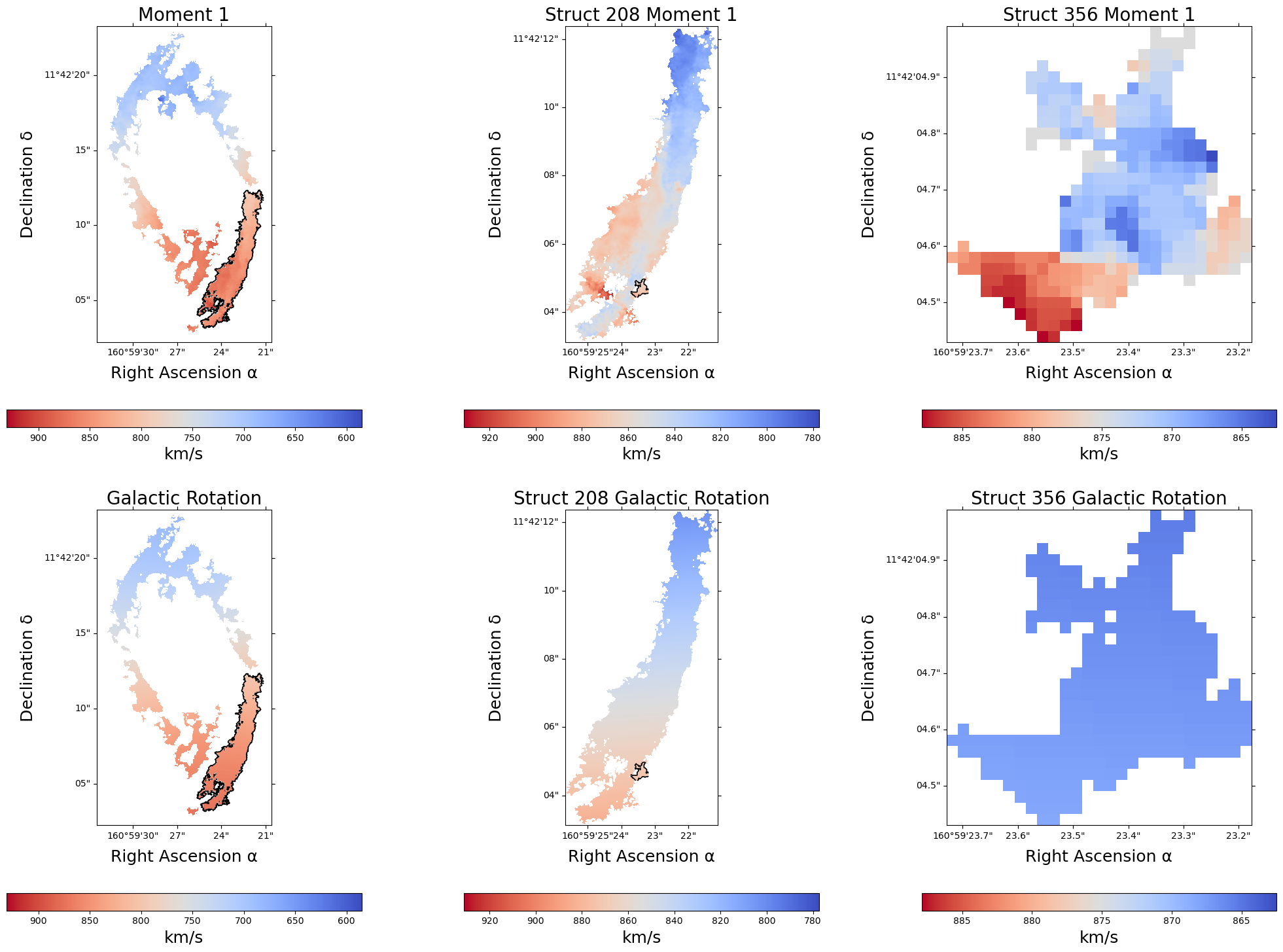}
\caption{
A comparison of our CO data to a pure galactic rotation model. The top row displays the CO moment 1 maps at different scales while the bottom row displays the galactic rotation model. We find that for larger structures, the galactic rotation model seems to describe the observed velocity gradients well, whereas in smaller structures, the model fails to capture the observed velocity gradient.
}
\label{fig:A1}
\end{figure*}

\begin{figure*}[htbp]
\centering
\includegraphics[width=1.0\linewidth]{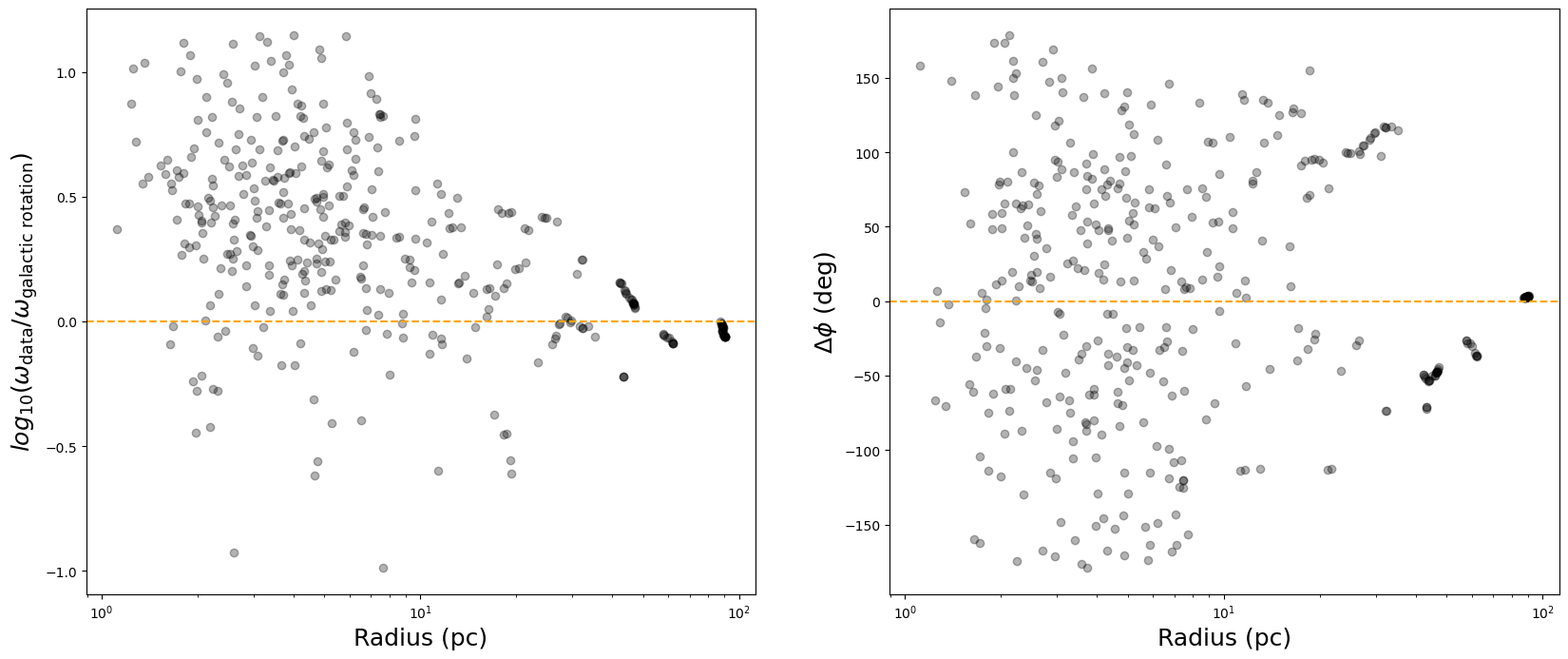}
\caption{
\textit{(Left)} Velocity gradient amplitude between the best-fit model and the galactic rotation model as a function of radius. We find that for larger structures ($>10\text{pc}$) the ratio approaches unity, which indicates that galactic rotation alone dominates the modeled ordered motion. For smaller structures, the galactic rotation model tends to underestimate the velocity gradient amplitude, suggesting substantial contribution from other sources of ordered motions.
\textit{(Right)} Difference in position angle between the best-fit model and the galactic rotation model. For the largest structures, the modeled ordered motion is well aligned with the expected direction from pure galactic rotation.
}
\label{fig:A2}
\end{figure*}

In \autoref{sec:results}, we examined various physical processes that can influence gas kinematics and dynamics over a wide range of spatial scales ($\sim$1--100~pc in terms of structure sizes). We showed that ordered and random motions tend to dominate the kinematics of larger and smaller gas structures, respectively. Modulo uncertainties on $\alpha_\mathrm{CO(3-2)}$, the strength of self-gravity of gas structures appear on par with the ordered motions at all scales probed by the data.

An important remaining question is the nature of the ``ordered motion'' captured by our analysis. In \autoref{sec:datamethod:vdisp}, we note that this ordered motion can include contributions from (differential) galactic rotation, non-circular streaming motions, and cloud spin or rotation. Differentiating these mechanisms can help us understand the cause of such motions (e.g., large-scale galactic potential, local gas self-gravity, stellar feedback, and/or cloud--cloud collisions) and correctly model the gas dynamics and stability in future studies.

As a first step of addressing this question, here we construct a model of pure galactic rotation and compare it to our measured ordered motion for individual gas structures. We use the arctan model of the galactic rotation curve fitted in \cite{Lang2020}:
\begin{equation}
V_{c} = V_{0}\arctan(\frac{r}{R_t})~,
\label{gal_rot}
\end{equation}
\noindent where $V_{0}=206.8\rm\;km/s$ is the maximum rotational velocity of NGC 3351 and $R_t=0.3\rm\;kpc$ is the relevant scale radius. We project this model onto the sky, taking into account the systemic velocity (778~km/s), inclination angle (45.1$^\circ$), and position angle (188.4$^\circ$) of NGC~3351. This generates a model for the pure galactic rotation across the entire CMZ. We then mask this model velocity field to match the CO emission footprint of the CMZ. As shown by \autoref{fig:A1}, the galactic rotation model matches the observed CO velocity field reasonably well for larger structures, but that is not the case for most smaller structures.

In order to quantify the strength of galactic rotation motion on the scales of individual gas structures in the dendrogram, we replicate our analysis in \autoref{sec:datamethod:vdisp}, using the galactic rotation model in place of the observed velocity field. That is, we fit a linear velocity gradient to the galactic rotation model within the footprint of each gas structure, using \autoref{eq:vmodel}. We then compare the amplitude and position angle of this new velocity gradient (capturing only galactic rotation) to that measured from the observed CO velocity field (from \autoref{sec:datamethod:vdisp}).

\autoref{fig:A2} shows the differences in the amplitudes (left panel) and position angles (right panel) between the best-fit velocity gradients on the galactic rotation model and the real CO data. We find that for small structures below $\sim$10~pc, both the amplitude and position angle show poor agreement. Furthermore, the velocity gradient amplitude measured from the real CO data tends to be larger than the galactic rotation model for most small structures. For larger structures, however, the amplitude and position angle show progressively better agreements.

Our interpretations of these results are as follows. For the largest structures in the dendrogram hierarchy, most of the ordered motion is attributable to galactic rotation alone, which is due to the large-scale galaxy potential. For smaller structures ($\lesssim$10~pc), the mismatch between amplitude and position angle implies other sources of ordered motion, such as non-circular streaming motion, cloud--cloud collisions, and/or gravitational collapse \citep[e.g.,][]{Ruffa2019,Vazquez-Semadeni2019}.

We also note that if large-scale turbulence is present, the velocity gradient across small gas structures can be partly attributed to this turbulence as well \citep[see e.g.,][]{Burkert2000}. Due to this effect, the correspondence between the ordered versus random motions that we measured and the actual turbulent versus non-turbulent motions in the gas is not straightforward. Strictly speaking, the only gas motion we can view as truly non-turbulent would be the galactic rotation motion. In this sense, the galactic rotation model would provide an alternative and arguably safer lower bound on the ordered motion velocity dispersion $\sigma_\mathrm{ord}$. This would not change our main conclusions for the largest gas structures but would leave much more ambiguity on the behaviors of smaller structures. We anticipate future works that model the local gravitational field and gas dynamics to shed more light on this matter.

\section{Conclusion}
\label{sec:conclusion}

We characterize the multi-scale molecular gas structure and dynamics using high-resolution ALMA \CO32\ data covering the CMZ of NGC~3351.
Our dendrogram analysis identifies 398 unique gas structures across a wide range of spatial scales (from a few parsecs to $\sim$100~pc, \autoref{fig:data}).
This approach allows us to examine the scale dependence of gas structural and dynamical properties and infer their physics drivers.
Our key findings are as follows:

\begin{enumerate}

\item We find a size--mass relation of $M_\mathrm{mol} \propto r_\mathrm{phys}^{2.54}$ and a size--linewidth relation of $\sigma_v \propto r_\mathrm{phys}^{0.58}$ (\autoref{fig:larson}). The latter slope is consistent with studies of CMZ gas in our galaxy and other nearby galaxies and hints at effects of large-scale ordered motions.

\item We model the ordered and random motions within each gas structure (\autoref{fig:struct}) and find that both components show increasing amplitude with spatial scale. Random motion appears to dominate the total velocity dispersion in smaller structures ($\lesssim$30~pc), while ordered motions from galactic rotation appears to dominate in larger structures (\autoref{fig:vdisp}, also see \autoref{sec:discuss}).

\item We find the crossing times for both types of motions and the gravitational free-fall time to be $\gtrsim$1--10~Myr among the identified structures (\autoref{fig:timescale}). 
These numbers suggest rapid evolution of gas structures in CMZs, due to either rapid collapse (when self-gravity dominates) or dissolution (when turbulence or shearing motion dominates). By comparing these timescales for structures of various sizes, we find that the ordered motion alone can counteract gas self-gravity on $\sim$10~pc scales. On smaller scales, random motions appear more than strong enough to provide support against self-gravity, though this result may be more affected by $\alpha_\mathrm{CO}$ uncertainties.

\end{enumerate}

Our work complements recent efforts examining molecular gas structural and dynamical properties as functions of spatial scale in massive star-forming galaxy disks \citep[e.g.,][]{Oakes2025,Xie2025}.
We anticipate ongoing and future studies to expand this into more diverse galactic environments and to more conclusively pinpoint the governing physics in different regimes.

\vspace{\baselineskip}
The authors would like to thank Xander~Jenkin and Adam~Leroy for helpful discussions.
JS acknowledges support by the National Aeronautics and Space Administration (NASA) through the NASA Hubble Fellowship grant HST-HF2-51544 awarded by the Space Telescope Science Institute (STScI), which is operated by the Association of Universities for Research in Astronomy, Inc., under contract NAS~5-26555.

This paper makes use of the following ALMA data: ADS/JAO.ALMA\# 2021.1.00059.S, 2022.1.00159.S. ALMA is a partnership of ESO (representing its member states), NSF (USA), and NINS (Japan), together with NRC (Canada), MOST and ASIAA (Taiwan), and KASI (Republic of Korea), in cooperation with the Republic of Chile. The Joint ALMA Observatory is operated by ESO, AUI/NRAO, and NAOJ. The National Radio Astronomy Observatory is a facility of NSF operated under cooperative agreement by Associated Universities, Inc.

We acknowledge the usage of the SAO/NASA Astrophysics Data System.

\facilities{ALMA}

\software{
\texttt{NumPy} \citep{NumPy_2020},
\texttt{SciPy} \citep{SciPy_2020},
\texttt{Matplotlib} \citep{Matplotlib_2007},
\texttt{Astropy} \citep{astropy2013,astropy2018,astropy2022},
\texttt{spectral-cube} \citep{spectralcube},
\texttt{Astrodendro} \citep{astrodendro},
\texttt{CARTA} \citep{CARTA_2.0.0},
\texttt{APLpy} \citep{APLpy_2012}.
}

\restartappendixnumbering
\twocolumngrid

%TC:endignore

\bibliographystyle{aasjournal}         % Use correct style
\bibliography{references}              % Replace with your .bib file

\end{document}